\documentstyle[11pt,paspconf,epsf]{article}
\def\edcomment#1{\iffalse\marginpar{\raggedright\sl#1\/}\else\relax\fi}
\reversemarginpar

\begin{document}
\title{The Chemical Enrichment History of Damped Lyman-alpha Galaxies}
\author{Limin Lu, Wallace L. W. Sargent, \& Thomas A. Barlow}
\affil{Caltech, 105-24, Pasadena, CA 91125 (email: ll@troyte.caltech.edu)}

\begin{abstract}
  Studies of damped Ly$\alpha$ absorption systems in quasar spectra
are yielding very interesting results regarding the chemical evolution
of these galaxies. We describe some preliminary results from
such a program.
\end{abstract}

\section{Introduction}

Damped Ly$\alpha$ absorption systems in quasar spectra  are generally 
believed to trace the absorption from interstellar gas in high-redshift 
galaxies, possibly from the disks or proto-disks of spirals (Wolfe 1988). 
They can be studied in the redshift range $0<z<5$ by combining UV and
optical observations.  The damped Ly$\alpha$ galaxies are particularly
suited for probing the chemical evolution of galaxies over a large
fraction of the Hubble time for several reasons:
(1) they are relatively common and easy to identify in quasar spectra, 
so building up a large sample is possible;
(2) given their large neutral hydrogen column densities 
(N(HI)$\sim 10^{20}-10^{22}$ cm$^{-2}$), most of the absorbing gas 
should be neutral so ionization corrections 
should be minimal (cf. Viegas 1995);
(3) the damped Ly$\alpha$ galaxies should be relatively representative of
galaxies at high redshifts since they are selected simply because they
happen to lie in front of background quasars.

  The first systematic investigation of the chemical evolution of
damped Ly$\alpha$ galaxies was conducted by Pettini and collaborators
(Pettini et al. 1994; 1995), who studied the Zn and Cr abundances 
in $\sim 20$ damped Ly$\alpha$ galaxies.
The advent of the 10-m Keck telescope allows us to carry out
similar investigations in a much more detailed fashion.
In this short contribution, we present some preliminary results from
such a program. Detailed analysis and discussion may be found in
Lu, Sargent, \& Barlow (1996; hereafter LSB96).

\section{Results}

    Figure 1 shows the abundance results so far obtained from our Keck
program, with the addition of selected measurements from published papers
where we believe the effect of line saturation has been treated properly.
We also correct (when applicable) the abundance measurements from previous 
papers for the set of new oscillator strengths compiled by 
Tripp, Lu, \& Savage (1995) so that all the measurements will be on the 
same footing.  References to the data used in constructing figure 1 
may be found in LSB96.

\eject
\centerline{}
\vskip 4.9in
\noindent{\sl Figure 1. (a) Age-metallicity relation for our sample of damped 
Ly$\alpha$ galaxies. The conversion from redshift to age is calculated for 
$q_0=0.5$ and $H_0=50$. (b)-(f) Abundance ratios of selected elements
found in damped Ly$\alpha$ galaxies. The notion [Fe/H] has the meaning
[Fe/H]=log(Fe/H)$_{damp}-$log(Fe/H)$_{\odot}$, and similarly for others.
Typical measurement errors of the abundances are 0.1 dex.}

\vskip 0.2in
\subsection{Age-Metallicity Relation}
   Figure 1(a) shows the age-metallicity relation for our sample of
damped Ly$\alpha$ galaxies (filled circles). The solid curve roughly
indicates the 
age-metallicity relation for disk stars in the Galactic solar 
neighborhood determined by Edvardsson et al. (1993). 
We note the following:

(1) The damped Ly$\alpha$ galaxies have Fe-metallicities ([Fe/H]) 
in the range of 1/10 to 1/300 solar, thus representing a population 
of very young galaxies at least in terms of the degree of chemical enrichment.

(2) The mean metallicity appears to increase with age, providing
direct evidence for the buildup of heavy elements in galaxies.
It may be significant that all the four galaxies with $z>3$ have
[Fe/H]$<-1.7$, while at $2<z<3$ at least some galaxies have achieved
much higher metallicities. This may signal an epoch of rapid star formation
in galaxies. We also note that the {\it intrinsic} trend of increasing 
metallicity with age would be stronger if Fe is somewhat depleted by dust
in these galaxies because the depletion should be the least for the
highest redshift galaxies (see section 2.2, however).

(3) Clearly the damped Ly$\alpha$ galaxies have much lower
metallicities than the Milky Way disk at any given time
in the past. This may bear significantly on the nature of the damped
Ly$\alpha$ galaxies. It was suggested initially (cf. Wolfe 1988) that
the damped Ly$\alpha$ absorbers may trace disks or proto-disks
of high-redshift spirals. But the low metallicities of damped Ly$\alpha$ 
galaxies cast some doubts on this interpretation. Timmes, Lauroesch, \& Truran 
(1995; also see Timmes 1995, this volume) suggested that
the abundance measurements of damped Ly$\alpha$ galaxies are consistent with
the chemical enrichment history of the Milky Way disk if the  
enrichment process in damped Ly$\alpha$ galaxies is delayed by $\sim$3 Gyrs
for some reason; this seems to place the Milky Way at a privileged position.
On the other hand, the metallicities found for our sample of
damped Ly$\alpha$ galaxies are very similar to those found for Galactic
halo stars and globular clusters, suggesting the possibility that damped
Ly$\alpha$ absorbers may represent a spheroidal component of high-redshift
galaxies.  This possibility has in fact already been suggested by
Lanzetta, Wolfe, \& Turnshek (1995) based on considerations of
gas consumptions in these galaxies.

\subsection{Abundance Ratios and Nucleosynthesis}

Panels (b)-(f) of figure 1 show the abundance ratios of various
elements in damped Ly$\alpha$ galaxies relative to their corresponding 
solar ratios.
Elemental abundance ratios, in principle, allow one to gain insight of
what kind of nucleosynthetic processes may be responsible for the
enrichment of the interstellar medium.  For example, the well-documented
overabundance of even-Z (Z=atomic number) $\alpha$-group elements relative 
to the Fe-peak elements in Galactic halo stars is believed to reflect the
nucleosynthetic products of massive stars through SN II explosions 
(cf. Wheeler, Sneden, \& Truran 1989).  It is interesting
that the observed abundance patterns of N/O, Si/Fe, Cr/Fe and Mn/Fe
in damped Ly$\alpha$ galaxies are all consistent with measurements
in Galactic halo stars (cf. Wheeler et al. 1989). In particular, 
we note that the observed N/O ratios
are not easily explained  with dust depletions because N and O are 
largely unaffected by dust in the Galactic ISM. The observed
Mn/Fe ratios are also difficult to explain with dust depletions because
in the Galactic ISM dust depletions cause the gas-phase Mn/Fe ratio to be 
higher than the solar
ratio, opposite to what is observed in damped Ly$\alpha$ galaxies. 
On the other hand, these ratios are easily understood in terms of
the odd-even effect (ie, the odd-Z elements generally show underabundances
relative to the even-Z elements of same nucleosynthetic origin at
low Fe metallicities) and the different nucleosynthetic origins of these
elements (cf. Wheeler et al. 1989).
These results strongly indicate that we have observed
these galaxies during the epoch when SN II are largely responsible
for the enrichment of the interstellar medium in these galaxies,
while low mass stars have not had enough time to evolve and to dump their
nucleosynthetic products into the interstellar medium through mass loss
and SN Ia. Thus the chemical enrichment process in these galaxies 
should not have proceeded more than 1 Gyr when they were observed.

  However, the observed Zn/Fe ratio in damped Ly$\alpha$ galaxies
is inconsistent with the above nucleosynthesis interpretation. 
In Galactic stars, Zn/Fe is found to be solar at all metallicities
(cf. Wheeler et al. 1989 and references therein). 
This difference may suggest 
that, while the observed relative abundance patterns in damped Ly$\alpha$
galaxies are dominated
by the effects of nucleosynthesis,  there is some dust
depletion effect on top of that. 
The presence of a small amount of dust in damped Ly$\alpha$ galaxies
has been claimed from the reddening
of the background quasars (cf. Pei, Fall, \& Bechtold 1991).
On the other hand, recent theoretical studies indicate
that Zn can be produced in large quantities in the neutrino driven
winds during SN II explosions (Hoffman et al. 1995; see
also Woosley 1995, this volume). Since SN II makes little Fe, 
a Zn/Fe overabundance may be possible in the ejecta of SN II.
The puzzle is then why Zn is observed to track 
Fe abundance in Galactic stars. If indeed the observed Zn/Fe overabundance
in damped Ly$\alpha$ galaxies is
caused by depletion of Fe onto dust grains, the [Fe/H] measurements in
figure 1(a) will underestimate the true Fe-metallicities by $\sim 0.5$ dex
(on average).

\section{Concluding Remarks}

   Damped Ly$\alpha$ galaxies provide the unprecedented opportunity to 
directly probe the chemical enrichment history of galaxies over
a large fraction of the Hubble time. Some intriguing results have
already emerged from the current study. However, many questions
remain, eg, why do  the damped Ly$\alpha$
galaxies have so low metallicities compared to the past history of
the Milky Way disk, and what are the implications? 
What is the significance of the large scatter
in the measured [Fe/H] at any give redshift? 
How big a role does dust play in modifying the observed abundances
and their interpretations?
Some of these issues will be addressed in more details in  LSB96.

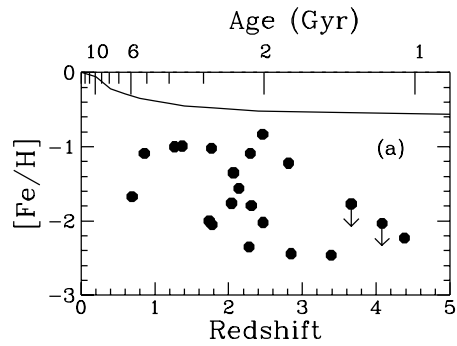
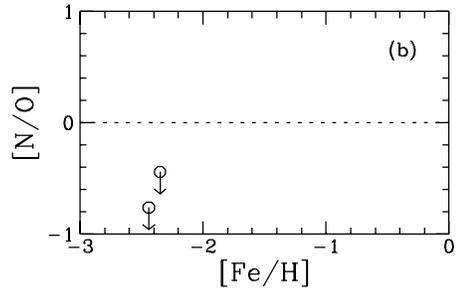
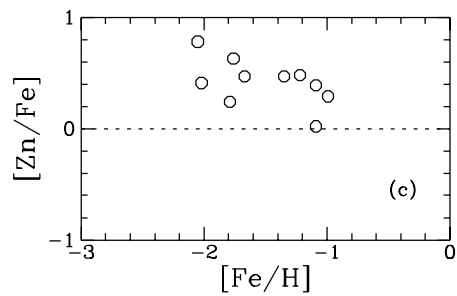
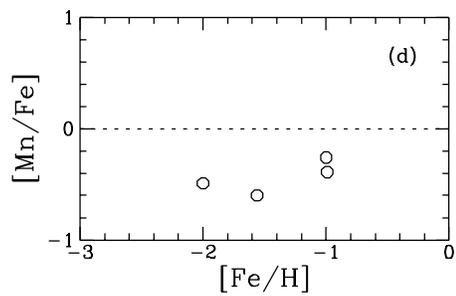
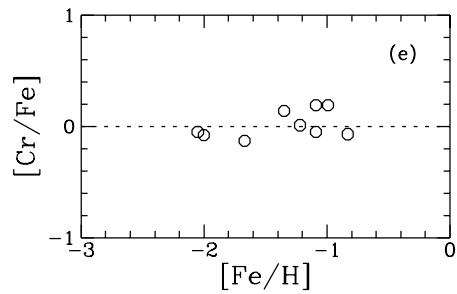
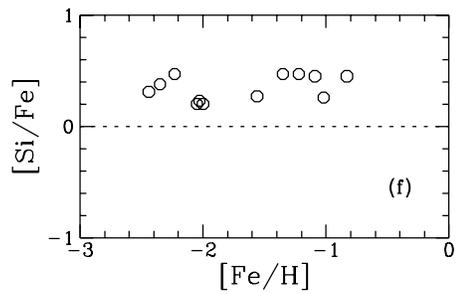
\end{document}